\documentclass[12pt]{article}
\usepackage{amssymb, amsmath}
\usepackage{amsthm}
\usepackage{mathabx}	

\newtheorem{lemma}{Lemma}
\newtheorem{proposition}{Proposition}
\newtheorem{theorem}{Theorem}
\newtheorem{corollary}{Corollary}

\usepackage{tikz}
\usetikzlibrary{matrix}

\usepackage[lined,boxed,linesnumbered,commentsnumbered]{algorithm2e}

\SetCommentSty{mycommfont}

\usepackage{mathptmx}       
\usepackage{helvet}         
\usepackage{courier}        
\usepackage{type1cm}        
\usepackage{verbatim}
\usepackage{makeidx}         
\usepackage{graphicx}        
\usepackage{multicol}        
\usepackage[bottom]{footmisc}

\usepackage[tight,footnotesize]{subfigure}

\usepackage{multicol}        
\usepackage[bottom]{footmisc}

\usepackage{url}

\author{
  Klaus Wehmuth\\
  National Laboratory for Scientific Computing (LNCC)\\
  Av. Get\'{u}lio Vargas, 333\\
  25651-075 -- Petr\'{o}polis, RJ -- Brazil\\
   \texttt{klaus@lncc.br}
  \and
     Eric Fleury\\
    ENS de Lyon / INRIA -- Universit\'{e} de Lyon\\
   UMR CNRS -- ENS Lyon -- UCB Lyon 1 -- INRIA 5668\\
   46, all\'{e}e d'Italie 69364 -- Lyon Cedex 07, France\\   
  \texttt{eric.fleury@inria.fr}
  \and
  Artur Ziviani\\
  National Laboratory for Scientific Computing (LNCC)\\
  Av. Get\'{u}lio Vargas, 333\\
  25651-075 -- Petr\'{o}polis, RJ -- Brazil\\
  \texttt{ziviani@lncc.br}
}
\title{A Unifying Model for Representing \\Time-Varying Graphs}
\date{}

\begin{document}
\maketitle

\begin{abstract}
Graph-based models form a fundamental aspect of data representation in Data Sciences and play a key role in modeling complex networked systems. In particular, recently there is an ever-increasing interest in modeling dynamic complex networks, i.e. networks in which the topological structure (nodes and edges) may vary over time. In this context, we propose a novel model for representing finite discrete Time-Varying Graphs~(TVGs), which are typically used to model dynamic complex networked systems. We analyze the data structures built from our proposed model and demonstrate that, for most practical cases, the asymptotic memory complexity of our model is in the order of the cardinality of the set of edges. 
Further, we show that our proposal is an unifying model that can represent several previous (classes of) models for dynamic networks found in the recent literature, which in general are unable to represent each other. In contrast to previous models, our proposal is also able to intrinsically model cyclic~(i.e. periodic) behavior in dynamic networks. These representation capabilities attest the expressive power of our proposed unifying model for TVGs. We thus believe our unifying model for TVGs is a step forward in the theoretical foundations for data analysis of complex networked systems.
\end{abstract}

\section{Introduction}
\label{sec:intro}

Data representation is a fundamental aspect in the field of Data Science~\cite{Dhar2013,Mattmann2013,jagadish2014}. In this context, graph models for representing complex networked systems find broad applicability in several different areas, ranging from techno-social systems~\cite{Vespignani2009} to computational systems biology~\cite{kitano2002}. In a typical graph model, nodes denote objects of the domain of interest (e.g., individuals or genes), and edges express interactions between these objects. Characteristics of objects and their interactions can be represented as node and edge properties, in a similar way to the attributes of relations. Indeed, advanced graph analytics is at the core of the new field of Network Science~\cite{Lewis2009,Kocarev2010,brandes2013}. 
Much of the utility of the graph abstraction, however, actually resides in the fact that it can represent relations between a set of objects as well as their connectivity properties, which derive from the notion of paths in a straightforward way without the need of further assumptions that are not explicit in the graph abstraction itself. 
Based upon graph representations, there is a lot of studies focused on investigating dynamic processes, such as random walks or information diffusion, over complex networks represented by graphs~\cite{Pastor-Satorras2000,Kempe2003,Barrat2008,Iliofotou2009a}. 

More recently, there is an ever-increasing interest in investigating not only the process dynamics on networks, but also the dynamics of networks, i.e. when the network structure (nodes and edges) may vary over time~\cite{ganguly2009,Figueiredo2012,Holme2013,Holme2015}. Indeed, Braha and Bar-Yam~\cite{Braha2006} indicate that new insights can be obtained from the dynamical behavior, including a dramatic time dependence of the role of nodes that is not apparent from static (time aggregated) analysis of node connectivity and network topology. Analyzing the dynamics of networks, however, brings a difficulty since the original graph abstraction was not created considering time relations between nodes. As a consequence, the need to extend the basic graph abstraction to include time relations between nodes arose, leading to many models for representing Time-Varying Graphs~(TVGs), also known in the literature as temporal or time-dependent networks~\cite{Hill2010, Xuan2003,Ferreira2004,Holme2005,Kostakos2009,Tang, Casteigts2012, Kim2012,Holme2012}.

Although recent TVG models appear extending the basic graph concept to include time relations~(Section~\ref{sec:rep} discusses related work), they are nonetheless not general enough to satisfy the needs of different complex networked systems and also in many cases rely on assumptions that are not explicitly expressed in the model. For instance, in models based on snapshots (i.e., a series of static graphs), as those in~\cite{Xuan2003,Ferreira2004}, it is implicitly assumed that a node in a given snapshot is connected to `itself' in the next snapshot, making it possible to extend the transitivity of edges over time. This assumption, however, is not made explicit in the snapshot model. Therefore, when analyzed without this implicit assumption, a snapshot model is a sequence of disconnected graphs and therefore no connectivity is possible between different time instants. The need to handle this assumption, which is not explicitly part of the model, brings difficulties since the structure of the model by itself is no longer sufficient to properly represent its behavior, making the semantics, applicability, interpretation, and analysis of such models more complex.

In this paper, we propose a new unifying model for representing finite discrete TVGs. Our proposed model is sufficiently general to capture the needs of distinct dynamic networks~\cite{Xuan2003,Ferreira2004,Holme2005,Kostakos2009,Tang, Casteigts2012, Kim2012,Holme2012,Alvarez-Hamelin2012,Sengul:2012,Guimaraes:2013,Martinet2014,Wang2014}, whereas not requiring any further assumption that is not already explicitly contained in the model itself. Further, our model aims at preserving the strictly discrete nature of the basic graph abstraction, while also allowing to properly represent time relations between nodes. The model we propose is based upon our recent work~\cite{Wehmuth2014a,Wehmuth2015}, which shows that structures such as multilayer and time-varying networks are in fact isomorphic to directed static graphs. Our proposal in this paper, however, is specifically tailored for use with TVGs.
Furthermore, we also demonstrate the unifying properties of our proposed model for representing TVGs by describing how it represents several previous (classes of) models for dynamic networks found in the recent literature, which in general are unable to represent each other. In contrast to previous TVG models, our proposal is able to intrinsically model cyclic~(i.e. periodic) behavior in complex dynamic networks. These representation features attest the expressive power of our proposed unifying model for TVGs. Finally,  we also demonstrate that, for most practical cases, the asymptotic memory complexity of our TVG model is determined by the cardinality of the set of edges in the model. We thus believe our unifying model for TVGs is a step forward in the theoretical foundations for data analysis of complex networked systems.\footnote{A preliminary version of this paper appears as a  technical report~\cite{Wehmuth2014}.}

This paper proceeds as follows. Section~\ref{sec:model_prop} introduces our proposed unifying model for representing TVGs and its main properties. Section~\ref{sec:tvg_rep} discusses data structures to properly represent TVGs using our model. In Section~\ref{sec:rep}, we show how our unifying model can be used to represent previous models for dynamic networks while these models in general are unable to represent each other. Finally, we conclude in Section~\ref{sec:conc}.

\section{Proposed model for representing TVGs}
\label{sec:model_prop}

The TVG model we introduce in this paper is a particular case of a MultiAspect Graph~(MAG)~\cite{Wehmuth2014a}, in which the vertices and time instants are the key features (i.e., aspects) to be represented by the model. 
A MAG is a structure capable of representing multilayer and time-varying networks, while also having the property of being isomorphic to a directed graph.
The MAG structural form is similar to the multilayer structure recently presented by~\cite{kivela2014}, since in both cases the proposed structure has a construction similar to an even uniform hypergraph associated with an adjacency concept similar to the one of simple directed graphs. 

Formally, a MAG can be defined as an object $H=(A,E)$, where $E$ is a set of edges and $A$ is a finite list of sets, each of which is called an aspect. In our case, for modeling a TVG, we have two aspects, namely vertices and time instants, i.e. $|A|=2$. For the sake of simplicity, this 2-aspect MAG can be regarded as representing
a TVG with an object $H = (V, E, T)$, where $V$ is the set of nodes, $T$ is the set of time instants, and $E \subseteq V \times T \times V \times T$ is the set of edges. 
As a matter of notation, we denote $V(H)$ as the set of all nodes in $H$, $E(H)$ the set of all edges in $H$, and $T(H)$ the set of all time instants in $H$. 

An edge $e \in E(H)$ is defined as an ordered quadruple $e = (u, t_a, v, t_b)$, where $u, v \in V(H)$ are the origin and destination nodes, 
respectively, while $t_a, t_b \in T(H)$ are the origin and destination time instants, respectively.
Therefore, $e = (u, t_a, v, t_b)$ should be understood as a directed edge from node $u$ at time $t_a$ to node $v$ at time $t_b$. If one needs to represent an undirected edge in the TVG, both $(u, t_a, v, t_b)$ and $(v, t_b, u, t_a)$ should be in $E(H)$. 

We also define four canonical projections, each projection mapping a dynamic edge into each one of its components:

\begin{eqnarray*}
\pi_1: E(H) & \to & V(H)\\
(u,t_a, v,t_b) & \mapsto & u,\\
& & \\
\pi_2: E(H) & \to & T(H)\\
(u,t_a, v,t_b) & \mapsto & t_a,\\
& & \\
\pi_3: E(H) & \to & V(H)\\
(u,t_a, v,t_b) & \mapsto & v,\\
& & \\
\pi_4: E(H) & \to & T(H)\\
(u,t_a, v,t_b) & \mapsto & t_b.\\
\end{eqnarray*}  

An edge $ e= (u, t_a, v, t_b)$ in our model may be classified into four classes depending on its temporal characteristic:
\begin{enumerate}
\item \emph{Spatial edges} connect two nodes at the same time instant, $e$ is in the form of  $e =(u, t_a, v, t_a)$, where $u \neq v$;
\item \emph{Temporal edges} connect the same node at two distinct time instants, $e$ is in the form of $e=(u, t_a, u, t_b)$, where $t_a \neq t_b$;
\item \emph{Mixed edges} connect distinct nodes at distinct time instants, $e$ is in the form of $e=(u, t_a, v, t_b)$, where $u \neq v$ and $t_a \neq t_b$;
\item \emph{Spatial-temporal self-loop edges} connect the same node at the same time instant, $e$ is in the form of $e=(u, t_a, u, t_a)$.
\end{enumerate}

The definition of a TVG $H$ is as general as possible and does not impose any order on the time set $T(H)$. One may, however, stick to the classical time notion and impose a total order on $T(H)$. In this context, where $T(H)$ has a linear order, both mixed or temporal dynamic edges $e = (u,t_a,v,t_b)$ can also be classified as progressive or regressive depending on the order of their temporal components. Dynamic edges that are originated at an earlier time instant and destined to a later time instant are progressive ($t_a < t_b$), whereas dynamic edges originated at a later time instant and destined to an earlier time instant are regressive ($t_a > t_b$). Regressive edges are particularly useful for creating cyclic TVGs, which in turn can be applied to model networks with a cyclic periodic behavior.

Further, we  define a \emph{temporal node} as an ordered pair $(u, t_a)$, where $u \in V(H)$ and $t_a \in T(H)$. 
The set $VT(H)$ of all temporal nodes in a TVG $H$ is given by the cartesian product of the set of nodes and the set of time instants, i.e. $VT(H) = V(H) \times T(H)$. 
As a notation note, a temporal node is represented by the ordered pair that defines it, e.g. $(u, t_a)$. 

The usage of the object $H=(V,E,T)$ to represent a TVG is formally described in the technical report~\cite{Wehmuth2014}. Therein, the representation of the TVG based on temporal nodes is proven to be isomorphic to a directed static graph. This is an important theoretical result since this allows the use of the isomorphic directed graph as a tool to analyze both the properties of a TVG and the behavior of dynamic processes over a TVG, as done is this work. This model is then shown to unify the representation of several previous (classes of) models for TVGs of the recent literature, which in general are unable to represent each other, as we show in further detail in Section~\ref{sec:rep}.

\section{Algebraic representations and structures}
\label{sec:tvg_rep}

\begin{figure}[h!]
 \centering
 \includegraphics[width=0.70\columnwidth,keepaspectratio=true]{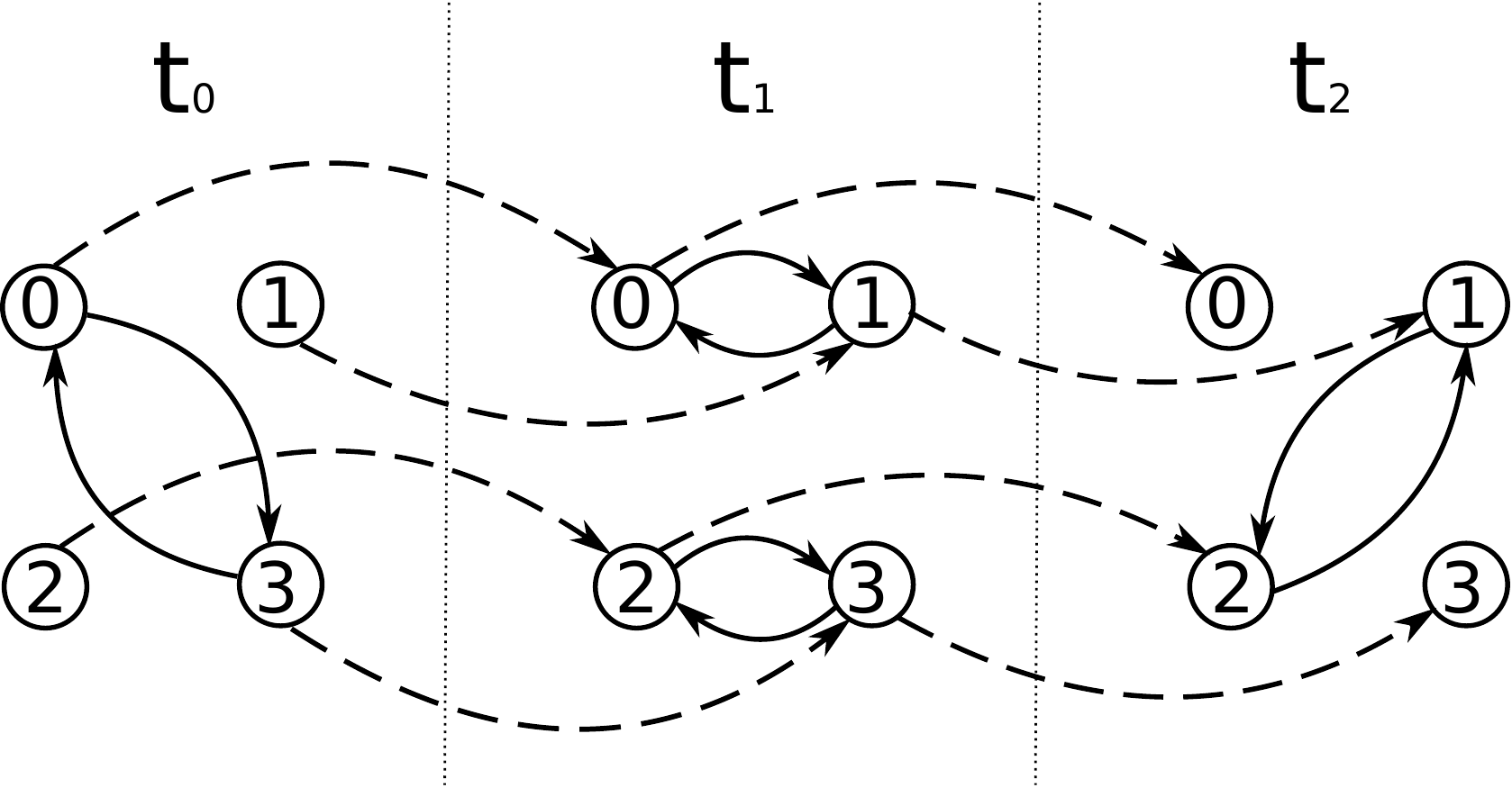}
 \caption{TVG $W$: Example for the algebraic representation.}
 \label{fig:exampleTVG}
\end{figure}

In this section, we discuss ways to properly represent a TVG using our proposed model. 
Similarly to static graphs, a TVG can be fully represented by an algebraic structure, like the MAG structure from which our TVG model is derived~\cite{Wehmuth2015}. 
We thus first present a TVG algebraic representation based on the adjacency tensor in Section~\ref{subsec:adj-tensor}.
In this work, we adopt matrix-based representations, in particular the adjacency matrix~(Section~\ref{subsec:mat-adj-tensor}).
We also present the TVG incidence matrix in Section~\ref{subsec:mat-ind-tensor}.
In order to illustrate such representations, we use the TVG $W$ presented in Figure~\ref{fig:exampleTVG}, where spatial edges are represented by solid arrows and temporal edges by dashed arrows. Finally, in Section~\ref{subsec:mem}, we analyze the memory complexity for storing a TVG using our model.

\subsection{TVG adjacency tensor}
\label{subsec:adj-tensor}
The TVG adjacency tensor follows from the adjacency matrix widely used to represent static graphs. However, since the dynamic edges used in the TVG are ordered quadruples, this representation has as a consequence to be done by means of a 4th order tensor.

We then define the adjacency tensor of a TVG $H$ as a 4th order tensor $\mathbb{A}(H)$ with dimension $|V(H)| \times |T(H)| \times |V(H)| \times |T(H)|$ that has an entry for every possible dynamic edge in~$H$. Each dynamic edge present in the TVG~$H$ is represented by a non-zero entry in the adjacency tensor~$\mathbb{A}(H)$, while all other entries  have a zero value. The non-zero entries represent the weight of the corresponding dynamic edge in the represented TVG. In the case of a unweighted TVG, the non-zero entries corresponding to the dynamic edges present in the TVG have value~1. The notation $\mathbb{A}(H)_{v, t_b}^{u, t_a}$ is used to identify the entry corresponding to the dynamic edge $(u,t_a, v,t_b)$ of the TVG~$H$.

As an example, the adjacency tensor of the TVG~$W$ depicted in Figure~\ref{fig:exampleTVG} has dimension $4 \times 3 \times 4 \times 3$, having a total of $144$ entries. For instance, the pair of dynamic spatial edges connecting node~$0$ at time~$t_0$ to node~$3$ at time~$t_0$ is represented by the entries $\mathbb{A}(W)_{3,t_0}^{0,t_0}$ and $\mathbb{A}(W)_{0,t_0}^{3,t_0}$, where both carry value~1 as the TVG~$W$ is unweighted.

Note that, even though the TVG adjacency tensor $\mathbb{A}(H)$ has dimension $|V(H_t)| \times |T(H_t)| \times |V(H_t)| \times |T(H_t)|$, only the entries corresponding to dynamic edges present in the TVG~$H$ have non-zero values. 

\subsection{TVG adjacency matrix}
\label{subsec:mat-adj-tensor}

Since every MAG has a directed static graph that is isomorphic to it~\cite{Wehmuth2014a}, the same holds for our TVG model, since it is a particular specialized case of a MAG. Consequently, it follows that the TVG can be represented by an adjacency matrix. 

\begin{figure}[h!]
 \centering
 \includegraphics[width=0.75\columnwidth,keepaspectratio=true]{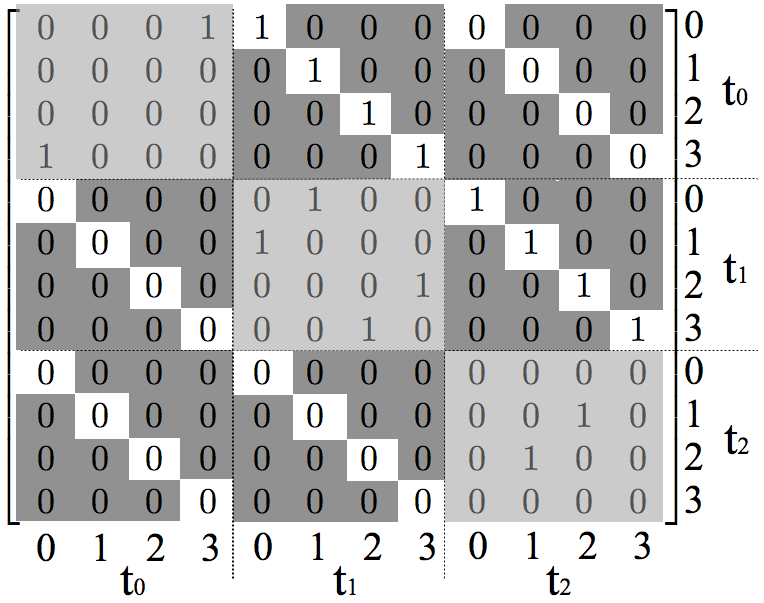}
 \caption{The matrix form of the adjacency tensor of TVG $W$.}
 \label{fig:MatAdj}
\end{figure}

In the more general environment represented by a MAG, a \emph{companion tuple} is used in order to properly identify and position each temporal node of the isomorphic graph in the adjacency matrix. Since the case we present in this work is restricted to MAGs with 2 aspects, it follows that the companion tuple is reduced to a pair, which in the first entry has the number of nodes and the second entry has the number of time instants.  For instance, considering the TVG example of Figure~\ref{fig:exampleTVG}, the companion tuple associated with its adjacency matrix is $(4,3)$, since there are $4$ nodes and $3$ time instants. The function of the companion tuple is only to ensure that the order by which the temporal nodes are placed in the adjacency matrix is the one shown in Figure~\ref{fig:MatAdj}. Since in the case where the number of aspects is restricted to $2$ this placement can be easily achieved, in this work we do not further mention the companion tuple.

To get the TVG adjacency matrix, we only need to consider that each temporal node $(u,t_a)$ can be thought of as a node in a static graph. This static graph has $|V| \times |T|$ nodes and, as a consequence, its adjacency matrix has $|V| \times |T| \times |V| \times |T| = |V|^2 \times |T|^2$ entries. Since the non-zero entries of this matrix correspond to the dynamic edges of the TVG, further analysis show that this matrix is usually sparse and can therefore be stored in an efficient way.

Figure~\ref{fig:MatAdj} shows the matrix representation obtained for the illustrative TVG~$W$ shown in Figure~\ref{fig:exampleTVG}. 
From Figure~\ref{fig:MatAdj}, we highlight that the matrix form of the TVG adjacency tensor has interesting structural properties. 
First, each one of the four nodes~(identified as~0, 1, 2, and 3) of the TVG $W$ clearly appears as a separate entity in each of the three time instants~($t_0$, $t_1$, and $t_2$) that compose the TVG~$W$. Second, the main block diagonal~(lightly shaded) contains the entries corresponding to the spatial edges at each time instant.
In these three blocks the entries corresponding to the spatial edges of the TVG carry value 1. Third, the unshaded entries at the off-diagonal blocks correspond to the temporal edges. The eight progressive temporal edges present at the TVG~$W$ are indicated by the value~1 on the first superior diagonal. Finally, the dark shaded entries are the ones that correspond to the mixed edges. Since no mixed edges are present in the example TVG~$W$, all these entries contain value~0. Further, 
we remark that the entries corresponding to progressive~(mixed and temporal) edges are above the main block diagonal, whereas the edges corresponding to regressive edges appear below the main block diagonal. All these structural properties derive from the order adopted for representing the nodes and time instants present in the TVG and can be readily verified in the matrix form in a quite convenient way. 

Note that the procedure used to obtain the matrix form of the adjacency tensor is the well-known matricization or unfolding of a tensor~\cite{Kolda2006}. In fact, since each entry of the matrix corresponds to an entry in the tensor, the process can be reversed, obtaining the corresponding tensor from its matrix form. 

\subsection{TVG incidence matrix}
\label{subsec:mat-ind-tensor}

The incidence matrix of a TVG is the incidence matrix corresponding to the directed static graph obtained though the temporal node representation presented in more detail in~\cite{Wehmuth2014}.
For instance, considering the TVG~$W$ shown in Figure~\ref{fig:exampleTVG}, there is a temporal edge connecting node~$0$ at time~$t_0$ to node~$0$ at time~$t_1$. Without loss of generality, we can label this edge as~$e_0$ and assign it the coordinate $0$ for the edge dimension of the incidence tensor. As a consequence, this edge is represented in the incidence tensor by setting $\mathbb{C}_{0,0}^0 = -1$ and $\mathbb{C}_{0,0}^1 = 1$.  Taking the temporal edge connecting node $0$ at time $t_1$ to node $0$ at time $t_2$ and labeling it as $e_1$, its representation on the incidence tensor is $\mathbb{C}_{0,1}^1 = -1$ and $\mathbb{C}_{0,1}^2 = 1$. By repeating this procedure for each edge in the TVG, the corresponding incidence tensor is created. Since the TVG~$W$ has $16$ edges, being~$8$ temporal edges and $8$ spatial edges,~$4$ nodes, and~$3$ time instants, the corresponding incidence tensor has a total of $4 \times 16 \times 3 = 192$ entries, from which $160$ carry value~$0$, $16$ carry value~$-1$, and $16$ value~$1$.  Note that the resulting matrix has $12$ rows (for 3 times and 4 nodes) and $16$ columns, one for each edge in the TVG, totalizing $12 \times 16 = 192$ entries. 
  
\begin{figure}[h!]
\centering
 \includegraphics[width=\columnwidth,keepaspectratio=true]{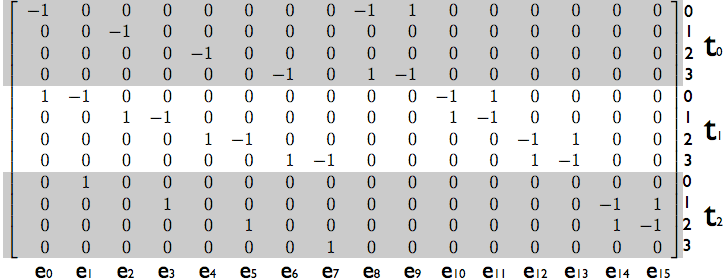}
 \caption{The matrix form of the incidence tensor of TVG $W$.}
 \label{fig:MatCr1}
\end{figure}

Figure~\ref{fig:MatCr1} shows the matrix representation of the incidence tensor corresponding to the TVG~$W$ shown in Figure~\ref{fig:exampleTVG}. In this representation, the eight temporal edges present in the TVG are labeled as edges $e_0$ to $e_7$ and the 8 spatial edges as edges $e_8$ to $e_{15}$. Considering this labeling, it is straightforward to verify the correspondence of the incidence tensor in matrix form to the TVG.

\subsection{Memory complexity for storing a TVG}
\label{subsec:mem}

In this subsection, we analyze the memory complexity for storing a TVG. This analysis is valid for the TVG representations based on adjacency and incidence as well as their corresponding algebraic structures, which have all been discussed in Sections~\ref{subsec:adj-tensor} to~\ref{subsec:mat-ind-tensor}.

In the general case, TVGs may have disconnected nodes or unused time instants.
A disconnected node is defined as a node that has no dynamic edge incident to it, i.e. a node with no connection at any time instant. In a similar way, an unused time instant in a TVG is defined as a time instant at which there is no dynamic edge originated from or destined to it.
In most practical cases, however, a TVG is expected to have none or very few disconnected nodes as well as a unused time instants as compared with the total number of nodes or time instants, respectively.

We show that when a TVG has no disconnected nodes and unused time instants, or if the number of disconnected nodes and unused time instants is significantly lower than the number of dynamic edges, then the amount of storage needed to represent a TVG is determined by the number of dynamic edges in the TVG. 
In other words, we show in this section that in most practical cases the memory complexity $M_C(H)$ for storing a given TVG~$H$ is $M_C(H) = \Theta(|E(H)|)$, where $E(H)$ is the set of dynamic edges in the TVG~$H$.

The first step to achieve this is to show that the set of connected nodes and the set of used time instants of a TVG can be recovered from the set of dynamic edges.
Lemma~\ref{lemma:red} shows this by using the dynamic edge definition and the canonical projections defined in Section~\ref{sec:model_prop}.

\begin{lemma}{In any given TVG $H$, the set of connected nodes and the set of used time instants can be recovered from the set $E(H)$ that contains the dynamic edges of $H$.}
\label{lemma:red}
\begin{proof}
Let $V_C(H)$ be the set of connected nodes on TVG $H$. We now show that $V_C(H)$ can be constructed from the set $E(H)$.
Since $V_C(H)$ contains only connected nodes, for every $u \in V_C(H)$ there is at least one dynamic edge incident to $u$. Let $e_u \in E(H)$ be a dynamic edge incident to node $u$. Then, either~$e_u$ is of the form $(u, \cdot, \cdot, \cdot)$ or of the form $(\cdot, \cdot, u, \cdot)$, and therefore, either $u = \pi_1(e_u)$ or $u = \pi_3(e_u)$. Therefore, we can write $V_C(H)$ as
\begin{equation*}
V_C(H) = \bigcup_{e \in E(H)} \left\{\pi_1(e), \pi_3(e)\right\}.
\end{equation*}

We now use a similar reasoning to recover the set of used time instants from the set $E(H)$.
Let $T_U(H)$ be the set of used time instants in $H$ and let $t_u$ be a used time instant. Then, there is at least one dynamic edge $e_u \in E(H)$ of the form $(\cdot, t_u, \cdot, \cdot)$ or $(\cdot, \cdot, \cdot, t_u)$ such that $t_u = \pi_2(e_u)$ or $t_u = \pi_4(e_u)$. Hence, we can write $T_U(H)$ as
\begin{equation*}
T_U(H) = \bigcup_{e \in E(H)} \left\{\pi_2(e), \pi_4(e)\right\}.
\end{equation*}
\end{proof}
\end{lemma}  

Since Lemma~\ref{lemma:red} shows that the both sets of connected nodes and used time instants in a given TVG~$H$ can be recovered from the set of dynamic edges~$E(H)$, we can conclude that such information is redundant and does not need to be stored to represent a TVG. Therefore, to represent any given TVG~$H$, it suffices to store the set of dynamic edges~$E(H)$ as well as the sets of disconnected nodes and unused time instants of the TVG. 

We now demonstrate  in Theorem~\ref{theo:mem} that when the sets of disconnected nodes and unused time instants of a TVG are significantly smaller than the set of dynamic edges, which is actually expected in most practical cases, then the asymptotic memory complexity of the TVG representation is determined by the cardinality of the set of dynamic edges.

\begin{theorem}{If the set of disconnected nodes and the set of unused time instants are smaller than the set of dynamic edges, then the memory complexity $M_C(H)$ for storing a given TVG~$H$ is determined by the size of the set of dynamic edges, i.e. $M_C(H) = \Theta(|E(H)|)$.}
\label{theo:mem}
\begin{proof}
Let $H$ be an arbitrary TVG, $E(H)$ its set of dynamic edges, $V_C(H)$ its set of connected nodes, $V_N(H)$ its set of disconnected nodes, $T_U(H)$ its set of used time instants, $T_N(H)$ its set of unused time instants, and $M_C(H)$ the memory complexity for storing the TVG $H$. Note that $V(H) = V_C(H) \cup V_N(H)$ and $T(H) = T_U(H) \cup T_N(H)$, whereas $V_C(H) \cap V_N(H) = \emptyset$ and $T_U(H) \cap T_N(H) = \emptyset$.

Since Lemma~\ref{lemma:red} shows that $V_C(H)$ and $T_U(H)$ can be recovered from $E(H)$, it follows that to store a representation of $H$ in memory, it suffices to store $E(H)$, $V_N(H)$, and $T_N(H)$. We now analyze the asymptotic bounds for the memory complexity for storing these sets, together with the assumption that $|V_N(H)| + |T_N(H)| < |E(H)|$.
\begin{itemize}
\item Lower bound for $M_C(H)$: 
Since for storing the set $E(H)$ it is necessary at least to store an ordered quadruple for each dynamic edge $e \in E(H)$, the memory needed is $c_1 \times |E(H)|$, where $c_1$ is an integer constant. To store the set $V_N(H)$ it is necessary to store all nodes $u \in V_N(H)$ and therefore the memory needed is $c_2 \times |V_N(H)|$, while to store the set $T_N(H)$ it is necessary to store all time instants $u \in T_N(H)$, leading to a memory need of $c_3 \times |T_N(H)|$. Therefore, the total memory need is at least $c_1 \times |E(H)| + c_2 \times |V_N(H)| + c_3 \times |T_N(H)|$, and $M_C(H) = \Omega(|E(H)| + |V_N(H)| + |T_N(H)|)$. Finally, from our assumption that $|V_N(H)| + |T_N(H)| < |E(H)|$, we conclude that $M_C(H) = \Omega(|E(H)|$).

\item Upper bound for $M_C(H)$: On the upper bound analysis,  we see that the information to be stored is at most the same three sets $E(H)$, $V_N(H)$, and $T_N(H)$, which had to be stored in the lower bound analysis. Therefore, we conclude that $M_C(H) = O(|E(H)|$).
\end{itemize}
Since $M_C(H) = \Omega(|E(H)|)$ and also $M_C(H) = O(|E(H)|)$, we finally conclude that $M_C(H) = \Theta(|E(H)|)$, if $|V_N(H)| + |T_N(H)| < |E(H)|$.
\end{proof}
\end{theorem}

\begin{corollary}{The complete expression of the memory complexity for a given TVG~$H$ is $M_C(H) = \Theta(|V_N(H)| + |T_N(H)| + |E(H)|)$.}
\begin{proof}
It follows from the proof of Theorem~\ref{theo:mem} that the formal and complete expression of the lower bound memory complexity for storing a given TVG~$H$ is $M_C(H) = \Omega(|V_N(H)| + |T_N(H)| + |E(H)|)$, where $V_N(H)$ its set of disconnected nodes, $T_N(H)$ its set of unused time instants, and $E(H)$ the set of dynamic edges, while the upper bound is $M_C(H) = \Omega(|V_N(H)| + |T_N(H)| + |E(H)|)$. We therefore conclude that complete form of the memory complexity is 
$M_C(H) = \Theta(|V_N(H)| + |T_N(H)| + |E(H)|)$.
\end{proof}
\end{corollary}

Since in most practical cases the amount of disconnected nodes and unused time instants is significantly smaller than the number of dynamic edges 
(i.e., $|V_N(H)| + |T_N(H)| \ll |E(H)|$), we can conclude that the expected amount of memory needed to store a representation of a given TVG~$H$ is determined by the size of the set of dynamic edges~$|E(H)|$, i.e. $M_C(H) = \Theta(|E(H)|)$ as states Theorem~\ref{theo:mem}. Furthermore, any given TVG~$H$ for most practical applications can typically be expected to be \emph{sparse}, i.e. the number of dynamic edges is significantly smaller than the squared number of temporal nodes~($|E(H)| \ll |VT(H)|^2$), thus allowing its storage in a compact form, similar to the compressed forms used for sparse matrices~\cite{templates}.

\section{A unifying TVG model}
\label{sec:rep}

In this section, we show that the TVG model we propose can be used to represent many previous models found in the literature. Further, these models are not always capable of representing each other and none of them has the same representation range of the unifying model we propose. 

In order to assert that all the considered models can be directly represented by our model, we show that each of these models is in fact a subset of what is representable in our model and can therefore be easily represented in our model, maintaining any previously obtained result as at least valid on a special case. 
To achieve this, we show that each of the studied models can be represented in our TVG model by using only a subset of the types of dynamic edges available on our model~(see Section~\ref{sec:model_prop}). We achieve this by presenting the general structure of the matrix form of the adjacency tensor of the TVG that represents the model under study. 
We establish a set of four nodes $V = \{0,1,2,3\}$ and a set of three time instants $T = \{t_0, t_1, t_2\}$ to be used to illustrate all of the following analyses.
This is done without loss of generality, since the same procedure could be applied to any set of nodes and time instants present in a TVG. 

To show the representation in our model of the different TVG models found in the literature, we group them into classes that can be represented in our unifying model in similar ways and analyze these classes in the remaining of this section.
 
\subsection{Models based on snapshots}
\label{subsec:snap}

Some snapshot models for TVGs adopt an aggregate graph and a sequence of successive state sub-graphs that represent the network in a discrete way as time passes. Some examples are the models proposed by  Ferreira~\cite{Ferreira2004} and Xuan~et~al.~\cite{Xuan2003}.
We also consider in this same class of snapshot models the models proposed by Holme~\cite{Holme2005} and Holme and Saram\"{a}ki~\cite{Holme2012}, where edges are represented as triples of the form~$(i,j,t)$ meaning the existence of a contact between nodes~$i$ and~$j$ at time~$t$. Still under this same class, we also consider the models proposed by 
Tang~et~al.~\cite{Tang, Tang2010, Tang2010a}, as well as all other models in which the TVG is proposed as a sequence of static graphs~(i.e., the snapshots), each of them representing the TVG at a given time instant. 

Snapshot models are widely used in the literature and in general give an intuitive and straightforward notion of TVGs. Nevertheless, the snapshot models also demand some assumptions to be made without having them explicitly constructed in the model. For instance, it is usually assumed in such snapshot models that the nodes have a sort of memory that allows the transitivity induced by the edges to propagate on each node over time. This means that a path can be constructed passing through a given node even if this node is disconnected from all others during a period of time, meaning that the node is capable of retaining the edge transitivity during the period of disconnection. Even though this behavior is straightforward and intuitively expected in many cases, it is not constructed within the model and has to be assumed as an additional (external) property of the model, resulting that this assumption has also to be incorporated into the algorithms used with those model and thus making these algorithms dependent of these external assumptions.

\begin{figure}[h!]
 \centering
 \includegraphics[width=0.75\columnwidth,keepaspectratio=true]{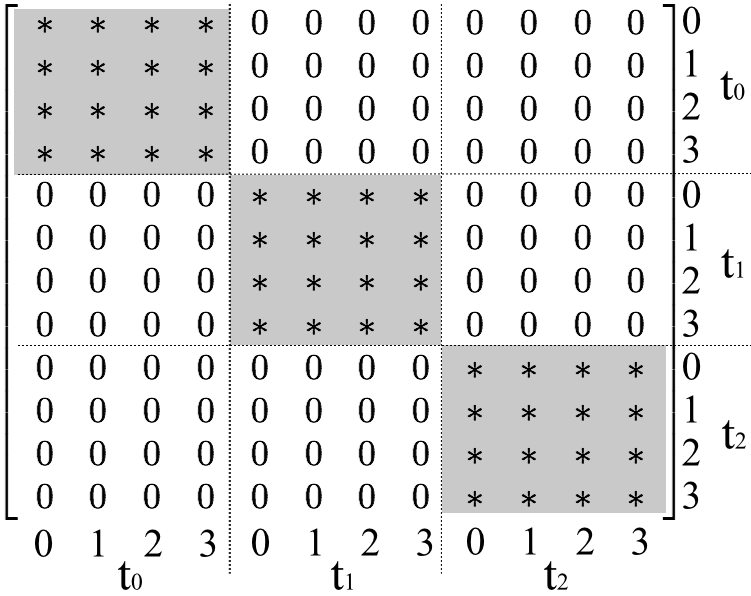}
 \caption{Snapshot models represented by our unifying TVG model.}
 \label{fig:MatAdjSnap}
\end{figure}

From the analysis of the snapshot models, we remark that clearly they only use edges connecting nodes in the same given time instant. This is consistent with the concept of spatial dynamic edges proposed in our model. In this way, all the TVGs constructed in this class of snapshot models can be represented in our model by using only spatial dynamic edges, given that all the necessary nodes and time instants are present in our model. Thus, a TVG with four nodes and three time instants in this class of snapshot model would be represented in our model by a TVG whose adjacency tensor in matrix form is as the one presented in Figure~\ref{fig:MatAdjSnap}. The entries containing ``*" may have non-zero values, indicating the potential presence of a spatial edges. Note that all the entries of this kind are located in the main block diagonal of the adjacency tensor in matrix form, which actually corresponds to the snapshots. 
 
In our general representation, we allow spatial and temporal self-loops (i.e. dynamic edges connecting a node to itself at the same time instant). This kind of edge is represented on the main diagonal of the matrix, just as it would be on an adjacency matrix of a static graph.

Any TVG of the snapshot class can be represented in this straightforward way in our proposed unifying model. Actually, snapshot models are in fact a sub-space of the representable TVGs in our model, making clear that this whole class of snapshot models is rather a subset of the representable TVGs in our model. Further, in snapshot models, each snapshot is formally disconnected from each other (although arguably implicitly connected), since the temporal connections are not explicitly constructed in the snapshot models.

The number of dynamic edges used to represent a TVG of this kind is the same as the number of edges present in the original snapshot-based representation. Therefore, the memory complexity for storing TVGs in our model is the same found on the original snapshot-based representation, which is compatible with the memory complexity discussed in Section~\ref{subsec:mem}. 

\subsection{Models based on continuous time intervals (CTI)}
\label{subsec:cti}

The class discussed in this subsection includes models that use a presence function defined over continuous time intervals~(i.e. $t \in \mathbb{R}^+$), such as the continuous time version of TVG proposed by Casteigts~et~al.~\cite{Casteigts2012}. We further assume that the presence function used in such models is constructed in such a way that every time interval (or their union) has a non-zero and finite measure. Although this assumption is not explicitly stated in the original paper, it is consistent with all examples and the reasoning present therein. It is also important to remind that the model we construct is a discrete version of the continuous time interval model that nonetheless retains all information present on the original model based on continuous time intervals.

\begin{figure}[h!]
 \centering
  \hspace{0.3cm}
 \subfigure[TVG model based on continuous time intervals]{
 \includegraphics[width=0.30\columnwidth]{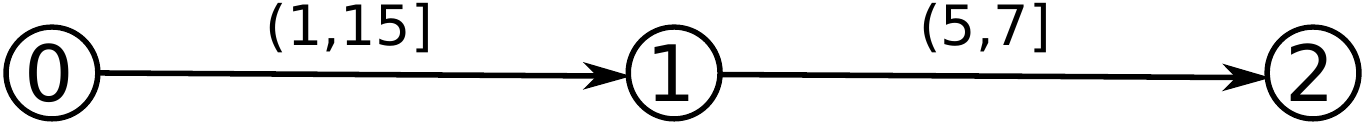}
 \label{fig:Int_1}}
 \vspace{0.5cm}
  \hspace{1.9cm}
 \subfigure[First representation: Use of mixed edges]{
 \includegraphics[width=0.45\columnwidth]{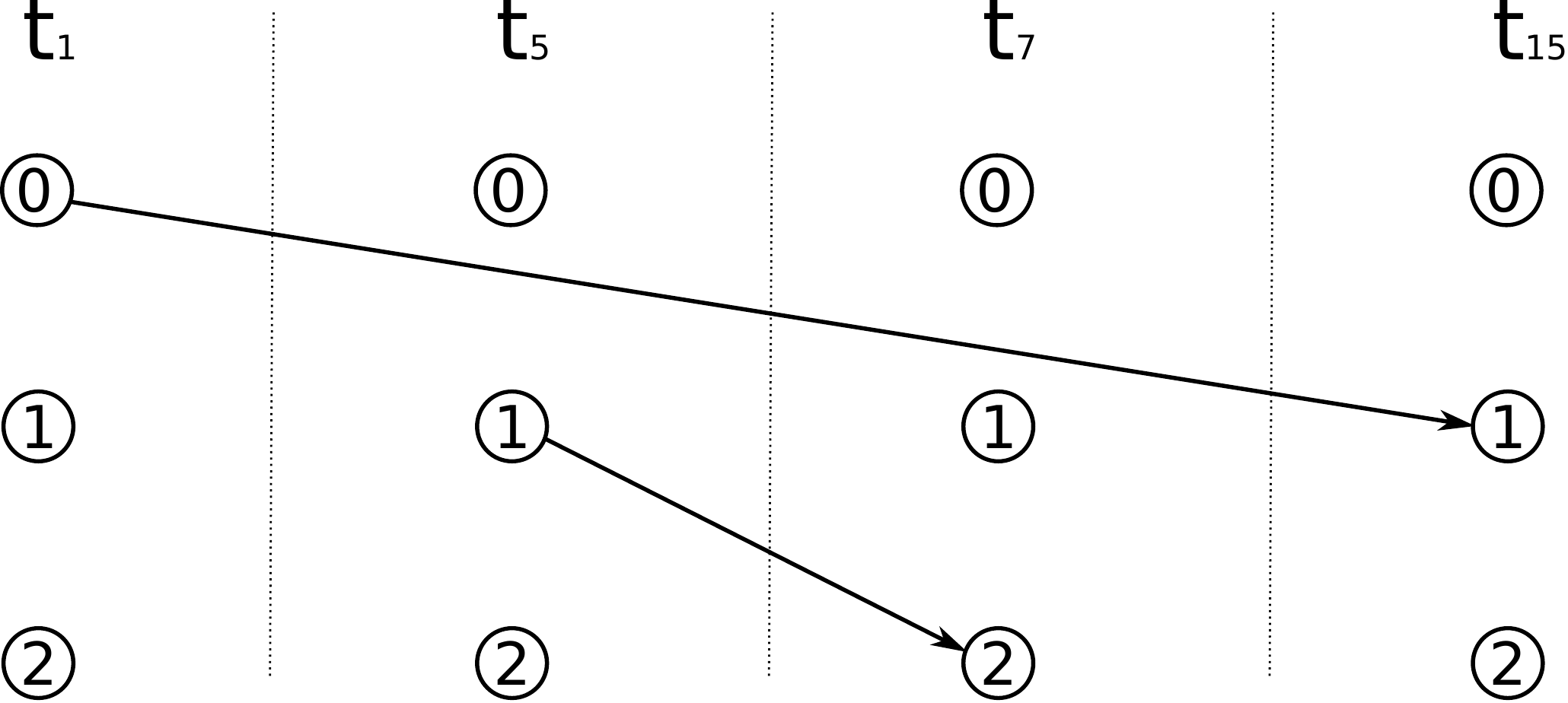}
 \label{fig:Int_2}}
 \vspace{0.3cm}
 \subfigure[Second representation: Snapshots]{
 \includegraphics[width=0.45\columnwidth]{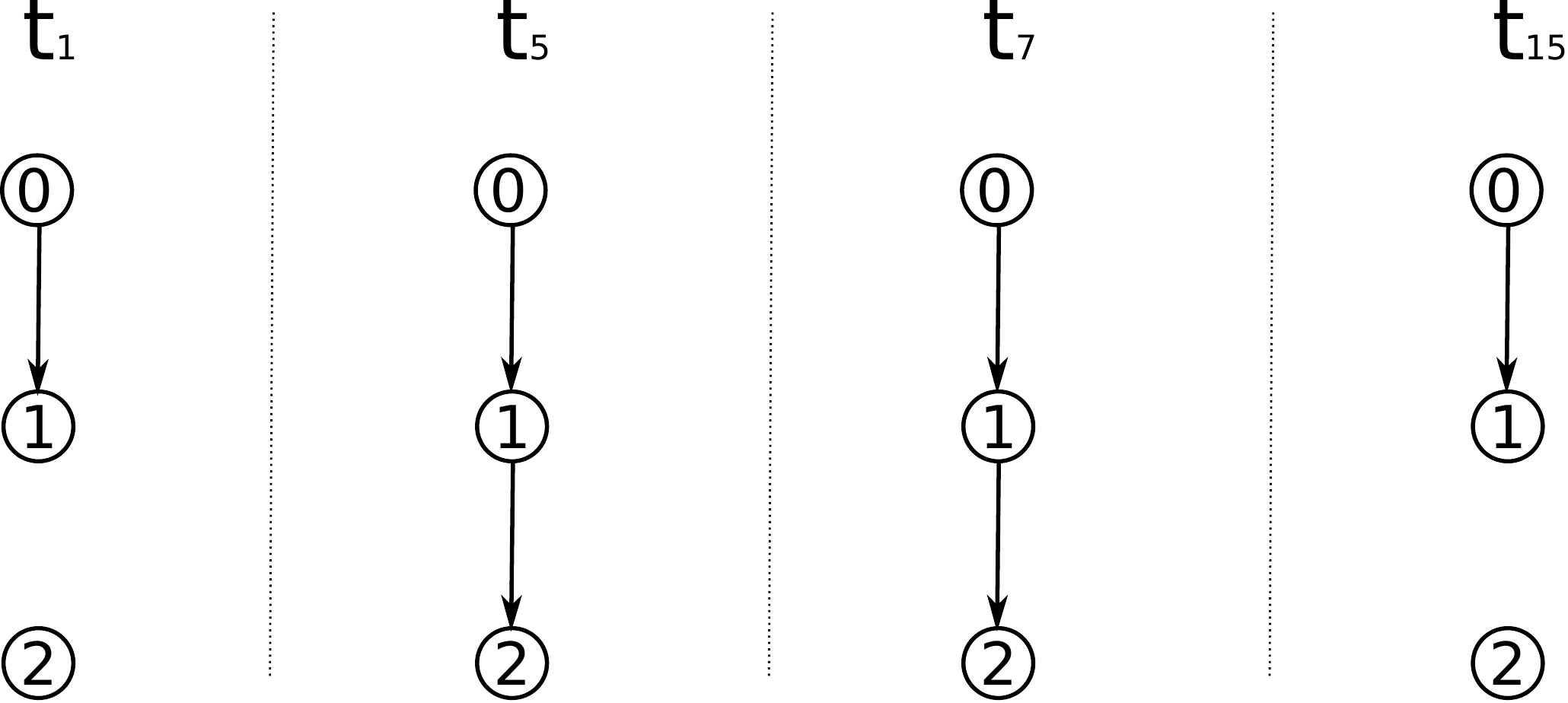}
 \label{fig:Int_3}}
  \vspace{0.3cm}
 \hspace{0.5cm}
 \subfigure[Third representation: Use of temporal and spatial edges]{
 \includegraphics[width=0.45\columnwidth]{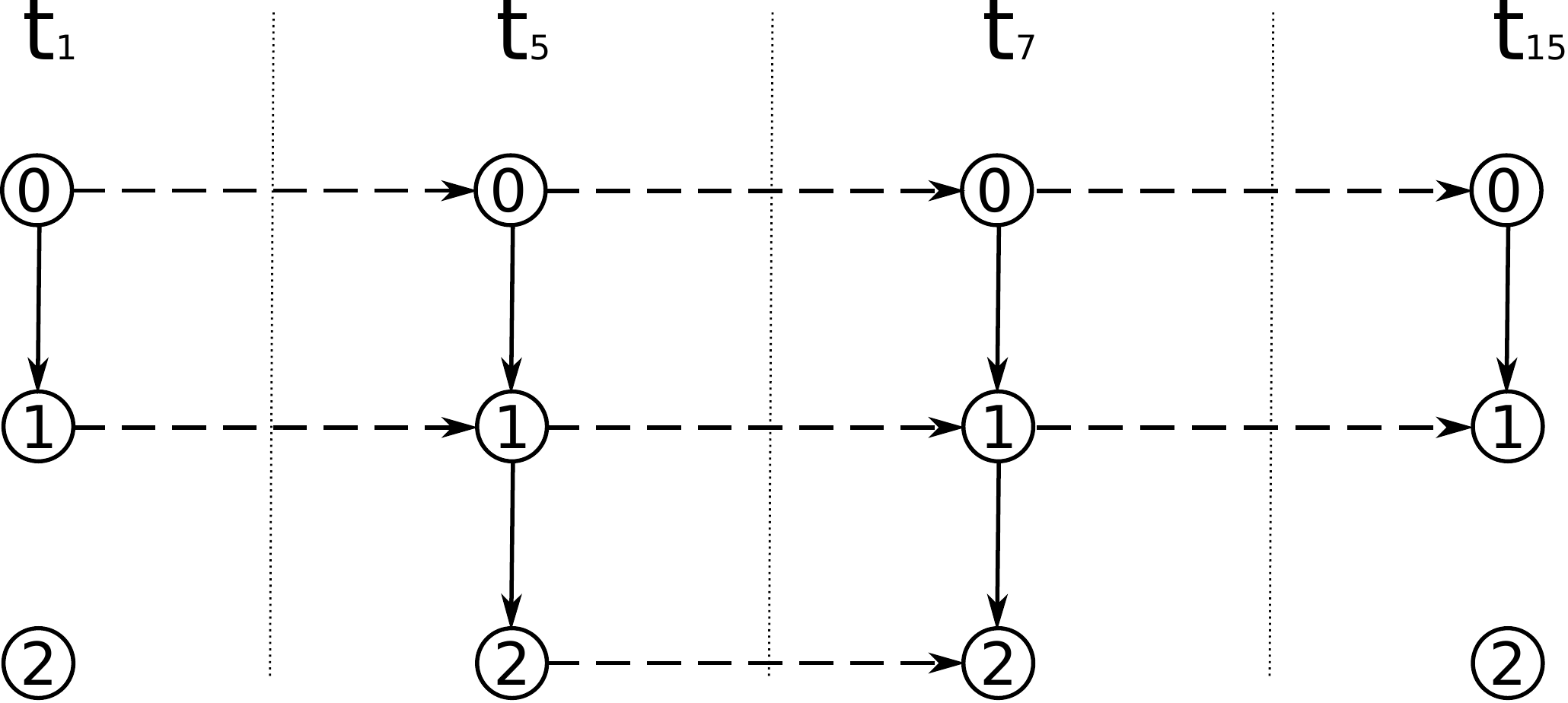}
 \label{fig:Int_4}}
  \vspace{0.3cm}
   \caption{Converting continuous time intervals into a discrete TVG.}
  \label{fig:Int}
 \end{figure}

In order to represent TVGs based on continuous time intervals in our unifying model, 
some representations are possible depending on the target application of the model that defines the semantics associated with an edge in the TVG.
An example with three possible representations is presented in Figure~\ref{fig:Int} and each of the these three representations is explained in the following.

If the semantic of a time interval $(t_a, t_b]$ associated with an edge $e$ between two nodes $u$ and $v$ is that the edge exists from the beginning of the interval (i.e. from time $t_a$) until the end of the interval (i.e. until time $t_b$), it is possible to represent the existence of such an edge by using a mixed dynamic edge. This is consistent with the original semantics, where the edge $e$ exists in the interval delimited by the mixed edge (i.e. from $t_a$ until $t_b$). A concrete example of this first representation is shown comparing Figures~\ref{fig:Int_1} and~\ref{fig:Int_2}. The edge between nodes $0$ and $1$ present at the time interval $(1, 15]$ shown in Figure~\ref{fig:Int_1} is represented by the mixed dynamic edge $(0,t_1, 1,t_{15})$ in Figure~\ref{fig:Int_2}. Similarly, the edge between nodes~$1$ and~$2$ at the time interval~$(5, 7]$ is represented by the mixed dynamic edge~$(1,t_5, 2,t_7)$.

In the case an edge is present at different time intervals, we represent this edge by having one mixed edge for each time interval.
Further, if the edge between nodes~$u$ and~$v$ is bidirectional, this edge is represented at each time interval by the pair of edges~$(u,t_a, v,t_b)$ and~$(v,t_a, u,t_b)$. Note that once the edges present in the TVG are defined for all time intervals, it is possible to construct the set~$T$ of time instants based on the dynamic edges present in the set~$E$. 
 From this, we conclude that the TVGs in this class modeled by continuous time intervals can be represented in our model using only progressive mixed dynamic edges.
 
\begin{figure}[h!]
 \centering
 \includegraphics[width=0.75\columnwidth,keepaspectratio=true]{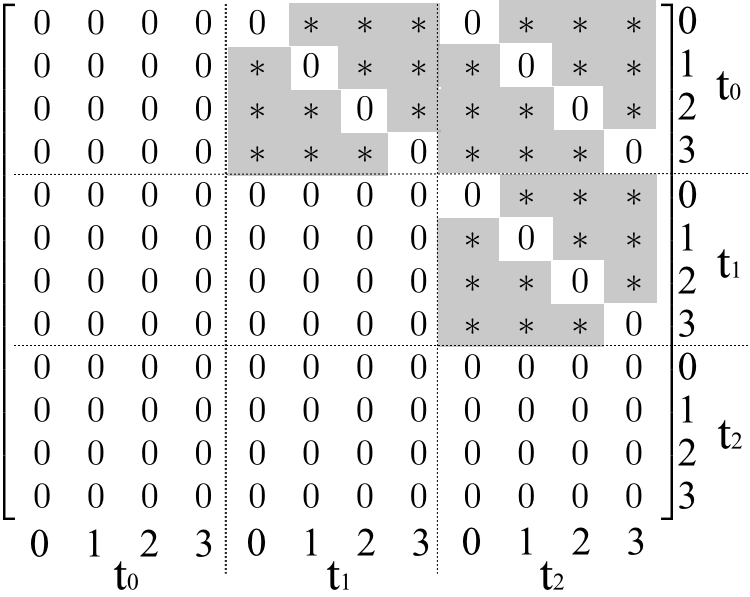}
 \caption{The first representation of TVGs based on continuous time intervals.}
 \label{fig:MatAdjIntMix}
\end{figure}

Figure~\ref{fig:MatAdjIntMix} shows the possible non-zero entries for a TVG with four nodes and three time instants based on the model of continuous time intervals using the first representation we are describing. Clearly from Figure~\ref{fig:MatAdjIntMix}, this kind of TVG is a subspace (and therefore a subset) of the TVGs representable by our proposed unifying model. Thus, we conclude that TVGs of this class are particular cases of the TVGs representable in our model. 
 
Even though this first representation by our model carries all information present in the original TVG based on continuous time intervals, it still relies on an assumption that is not explicitly present in this first representation. Namely, this assumption concerns that the time instants present on each mixed dynamic edge represent the time intervals on which the edge exists. Note that the same happens in the original model based on continuous time intervals. 
For example, using this assumption, a path between nodes~$0$ and~$2$ exists at any given time instant during the interval between~$t_5$ and~$t_7$.
To understand that this assumption is made in a manner that does not directly follows as a property of the structure of the TVG, notice that it makes the transitivity of the relation implied by an edge more difficult and cumbersome to determine. This happens because an external processing is needed to verify the presence of the edge~(either by the time intervals indicated on the mixed dynamic edge or by the set of intervals on the continuous time model) in order to be able to establish its transitivity to another edge. To establish the existence of a path connecting two nodes, it is necessary to compute the intersections of the time intervals on which each edge exists. The resulting path then exists on the time interval obtained by this intersection.
Therefore, the existence of a path connecting nodes in this kind of representation can be non-intuitive and computationally expensive to determine, thus impacting considerations about connectivity, reachability, communicability, and any other property derived from the transitivity of edges in a TVG. 

A second representation (Figure~\ref{fig:Int_3}) comes from the realization that the continuous time intervals model is in fact a continuos form of the snapshot model, in which edges are present at infinite (uncountable) time instants. Therefore, a natural way to represent the CTI model in a discrete form is to have the time instants set $T$ formed by the beginning and end instants of each interval present on the CTI model, and then placing the corresponding edge at each time instant for which an edge is present in the CTI model. The result of this is a snapshot representation of the CTI model, as shown in Figure~\ref{fig:Int_3}, where the edge between nodes $0$ and $1$ is present at instants $t_0, t_5, t_7$, and $t_{15}$, while the edge between nodes $1$ and $2$ is present at instants $t_5$ and $t_7$. This representation has the same characteristics we highlighted for the snapshot class of models. Although the TVG connectivity can be readily determined at each time instant, the snapshots (i.e. time instants) are formally disconnected from each other. Hence, if any sort of connectivity between time instants is to be considered, it has to be stated as an additional assumption, which is not directly encoded in the representation. 

\begin{figure}[h!]
 \centering
 \includegraphics[width=0.65\columnwidth,keepaspectratio=true]{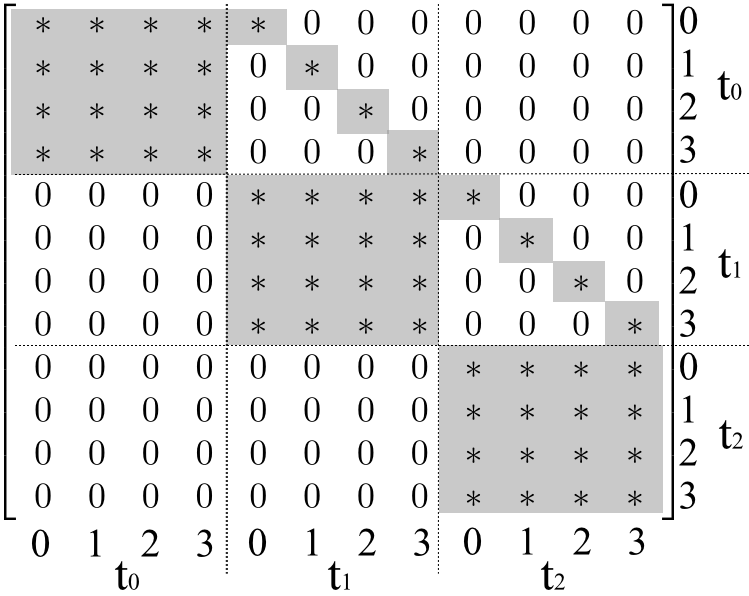}
 \caption{The third representation of TVGs based on continuous time intervals.}
 \label{fig:MatAdjIntDec}
\end{figure}

A third representation (Figure~\ref{fig:Int_4}) can be derived from the second one (snapshots), by formally placing temporal dynamic edges where the connectivity between successive time instants is desired. 
In our example, this leads to the representation shown in Figure~\ref{fig:Int_4}, where temporal nodes are used to connect nodes $0$ and $1$ between time instants $t_0, t_5, t_7$, and $t_{15}$ and to connect nodes $1$ and $2$ from time instant $t_5$ to $t_7$. This reflects the fact that in the original CTI model the edge $(0,1)$ was present at the interval $(t_0, t_{15}]$, while the edge $(1,2)$ was present at the interval $(t_5, t_7]$.  Figure~\ref{fig:MatAdjIntDec} shows the possible non-zero entries on the matrix form of the adjacency tensor of a TVG of this class using this third representation.

In order to determine the memory complexity of representing a TVG based on continuous time intervals in our proposed unifying model, we state the following proposition, based on the third representation, presented in Figure~\ref{fig:Int_4}, which uses the largest number of dynamic edges:

\begin{proposition}{A TVG based on continuous time intervals with $n$ nodes, $m$ edges, and $\eta$ continuous time intervals defining these edges can be represented in our model with $O(\eta^2)$ dynamic edges.}
\label{prop:cost}
\begin{proof}

For each time interval in the continuous time representation, a mixed dynamic edge is created on our model. Since each mixed edge requires two time instants to be defined, it follows that the representation in our model requires at most $2 \eta$ time instants. Therefore, in our model we have $|T| = 2\eta$ in the worst case.

To decompose a mixed dynamic edge into spatial and temporal dynamic edges, at most $|T|$ spatial dynamic edges are needed to connect the two nodes on the mixed edge in all possible time instants. Further, at most $2(|T| -1)$ temporal dynamic edges are needed to connect the nodes over all possible time instants. Therefore, to fully decompose a mixed edge, at most $2(|T|-1) + |T| = 3|T| - 2$ dynamic edges are needed. 

Hence, as $\eta$ mixed dynamic edges are needed in the representation, to fully decompose them into spatial and temporal edges, at most $\eta(3 |T| -2)$ dynamic edges are needed. Expanding and substituting $|T| = 2 \eta$ for the worst case, we have
\begin{eqnarray*}
\eta(3|T| -2) & = \eta(6 \eta - 2) \\
	           & = 6\eta^2 - 2\eta .\\
\end{eqnarray*}
We therefore conclude that the number of dynamic edges needed is $O(\eta^2)$.
\end{proof}
\end{proposition}

From this, we further conclude that, for a TVG with $\eta$ continuous time intervals, in our model we have $|E| = O(\eta^2)$. Therefore, the memory complexity of the representation based on continuous time instants is $O(\eta^2)$, which is compatible with the memory complexity discussed in Section~\ref{subsec:mem}.

\subsection{Models based on spatial and temporal edges (STE)}
\label{subsec:ste}

\begin{figure}[h!]
 \centering
 \includegraphics[width=0.75\columnwidth,keepaspectratio=true]{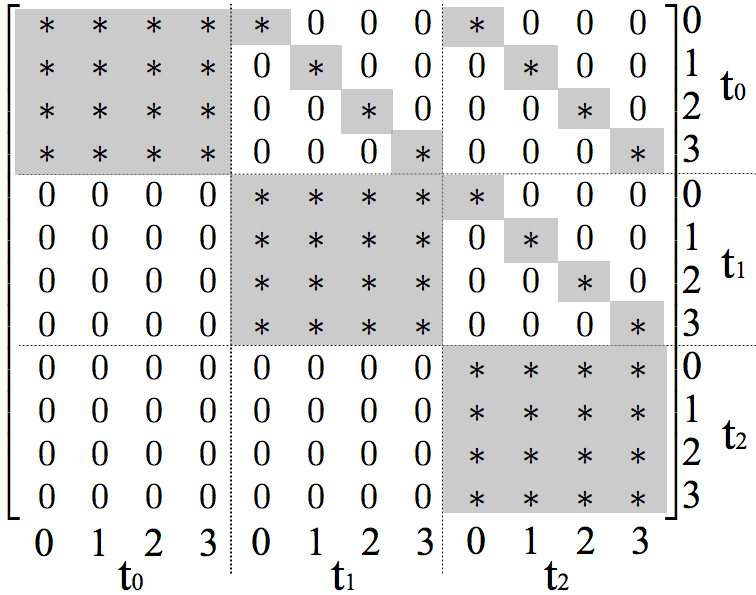}
 \caption{Representation of TVGs based on spatial and temporal edges.}
 \label{fig:MatAdjKostakos}
\end{figure}

Some models like the one proposed by Kostakos~\cite{Kostakos2009} are based on the idea that a class of links represent instantaneous iterations between distinct nodes while other class represent a waiting state of a given node. These concepts are formalized in our proposed unifying model by spatial and temporal dynamic edges. In our model, these dynamic edges are fully formalized and can be used to make a unambiguous representation of this kind of TVG.
Figure~\ref{fig:MatAdjKostakos} depicts the entries that may be non-zero on the matrix form of the adjacency tensor of a TVG of this class having four nodes and three time instants. It is again clear that this class of TVGs can be thus represented as a subset of the TVGs that can be represented by the unifying model we propose.

\subsection{Models based on temporal and mixed edges (TME)}
\label{subsec:tme}

\begin{figure}[h!]
 \centering
 \includegraphics[width=0.75\columnwidth,keepaspectratio=true]{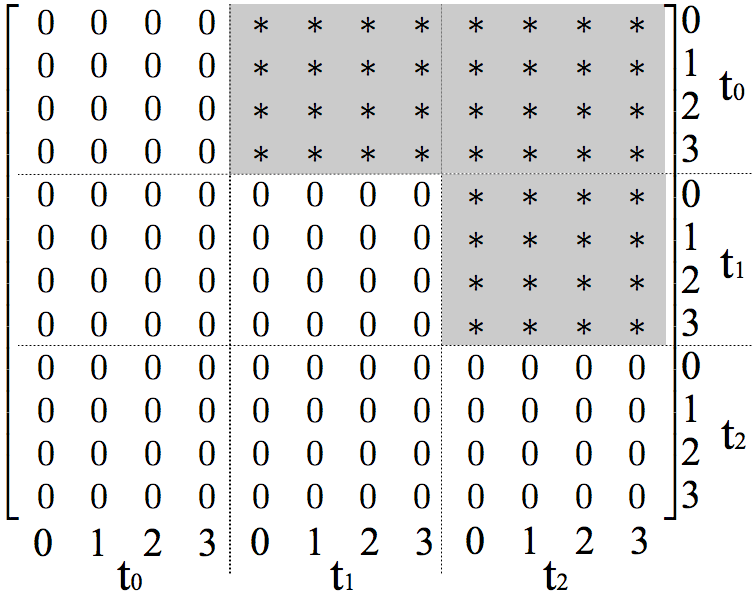}
 \caption{Representation of TVGs based on temporal and mixed edges.}
 \label{fig:MatAdjKim}
\end{figure}

Some works found in the recent literature, such as the one by Kim and Anderson~\cite{Kim2012}, loosely suggest the use of edges connecting nodes at different time instants. TVGs of this class can be represented in our model using only temporal and mixed dynamic edges. Figure~\ref{fig:MatAdjKim} shows the matrix form of the adjacency tensor of a TVG of this class. It can be seen that this is also a particular case of the TVGs that can be represented using our unifying model.

\subsection{Overview on the unifying representation of previous models}

Table~\ref{tab:rep_map} shows a representation map that indicates if a (class of) TVG model(s) is able to represent another (class of) TVG model(s). We compare the unifying 
representation model we propose with the models based on (i)~snapshots, (ii)~continuous time intervals~(CTI), (iii)~spatial and temporal edges~(STE), and (iv)~temporal and mixed edges~(TME), which have been presented in Sections~\ref{subsec:snap} to~\ref{subsec:tme}, respectively.

Snapshot models can only represent TVGs of the continuous time intervals~(CTI) class, since they are in fact a continuos time version of the snapshot model. Direct representation of  other models by snapshots is not possible because the snapshot model lacks the notion of temporal edges. 
Models based on continuous time intervals~(CTI) and spatial and temporal edges~(STE) can represent snapshot models because they support the notion of spatial edges as well as they are mutually able to represent each other. Models based on temporal and mixed edges~(TME) are able to represent models based on continuous time intervals~(CTI) in a discretized way. Remark that the converse is not true, i.e. models based on CTI are unable to represent models based on TME as they lack the notion of mixed edges. Furthermore, note that models based on TME are unable to represent in all cases models with spatial edges, such as the ones based on snapshots and STE. This is because, although a mixed edge could be seen as a composition of a spatial and a temporal edge, a mixed edge is unable to represent a single spatial edge, thus preventing the representation of these cases. 

Overall, we remark that considering the previous classes of TVG models, none is able to represent all others. This basically happens because the kind of edges present in one model not necessarily can be transformed into the kind of edges present in another model or class of models. In contrast, we have shown along this section that all classes of previous TVG models we consider can be represented in the unifying model we propose, whereas these previous TVG models not necessarily are able to represent each other, as shown in Table~\ref{tab:rep_map}.  Additionally, none of the previous TVG models makes use of regressive edges, which could be used in our proposal to intrinsically model cyclic~(i.e. periodic) behavior in dynamic networks. Therefore, we highlight that none of the previous analyzed models have all the representation capabilities and thus the same expressive power of our proposed unifying model for TVGs.

\setlength{\tabcolsep}{4pt}

\begin{table}
\centering
\caption{Representation map between TVG models: An entry is checked if the (class of) model(s) in the row is able to represent the (class of) model(s) in the column.}
\label{tab:rep_map}
\begin{tabular}{l c c c c c c c c c}
\hline \hline
							& Snapshots     & 	& CTI 				& 	&	STE		 		& 	& TME			    & 	& Unifying model\\	
\hline \hline
Snapshots			& \checkmark 	& 	&  \checkmark 					& 	& 						& 	& 						& 	&\\
CTI						& \checkmark	& 	& \checkmark	& 	& \checkmark			& 	& 						& 	&\\
STE						& \checkmark	& 	& \checkmark	& 	& \checkmark	& 	& 						& 	&\\
TME						& 						& 	& \checkmark	& 	&					 	& 	& \checkmark   & 		&\\
Unifying model		& 	\checkmark	& 	& \checkmark	& 	& \checkmark	& 	& \checkmark   & 		&\checkmark\\
\hline \hline
\end{tabular}
\end{table}

\setlength{\tabcolsep}{6pt}

\section{Summary and outlook}
\label{sec:conc}

In this paper, we have proposed a novel model for representing finite discrete Time-Varying Graphs~(TVGs). We have shown that our model is simple, yet flexible and efficient for the representation and modeling of dynamic networks. The proposed model preserves the discrete nature of the basic graph abstraction and has algebraic representations similar to the ones used on a regular graph. Moreover, we have also shown that our unifying model has enough expressive power to represent several previous (classes of) models for dynamic networks found in the recent literature, which in general are unable to represent each other, and also to intrinsically represent cyclic (i.e. periodic) behavior of dynamic networks. 
To further illustrate the flexibility of our model, we remark that our model can also be used to represent time schedules using a TVG. For example, in this case, the nodes can be thought of as locations and mixed edges as the amount of time taken to move between locations. Such representation can model scheduled arrivals and departures in transportation systems allowing, for instance, the evaluation whether a connection is feasible or the delivery time for logistics management. 

We have analyzed the proposed model proving that if the TVG nodes can be considered as independent entities at each time instant, the analyzed TVG is isomorphic to a directed static graph. 
This basic theoretical result has provided the ground for achieving other theoretical results that show that some properties of the analyzed TVG can be inferred from the temporal node representation of that TVG. This is an important set of theoretical results because this allows the use of the isomorphic directed graph as a tool to analyze both the properties of a TVG and the behavior of dynamic processes over a TVG. We have also demonstrated that, for most practical cases, the asymptotic memory complexity of our TVG model is determined by the cardinality of the set of edges. Further, from the basic definition of the model we proposed for representing TVGs, we have derived some basic properties such as communicability and connectivity using only properties that follow directly from the underlying structure of the model, such as the transitivity of the relation induced by the dynamic edges. As a consequence, in contrast to previous works, our model can be used without the need of external assumptions, meaning that all properties are derived from explicit properties of the TVG, in the same way that happens with static graphs. 

As future work, we intend to apply our proposed TVG model in the analysis of different dynamic complex networks as well as of properties over such dynamic networks. An example of such analysis is the recent work by Costa et~al.~\cite{Costa2015}. In that work, authors analyze the concept of time centrality for information diffusion in dynamic complex networks using the TVG model we propose in this paper, thus investigating the relative importance of different time instants to start a information diffusion process.

\section*{Acknowledgment}

The cooperation among authors has been supported by STIC-AmSud. K. Wehmuth and A. Ziviani acknowledge that their work was partially funded by the Brazilian funding agencies CAPES, CNPq, and FAPERJ as well as the Brazilian Ministry of Science, Technology, and Innovation~(MCTI). E. Fleury acknowledges the support by MILYON.

\end{document}